\documentstyle[11pt]{article}
\newcommand{\be}{\begin{equation}}
\newcommand{\ee}{\end{equation}}
\newcommand{\bea}{\begin{eqnarray}}
\newcommand{\eea}{\end{eqnarray}}
\newcommand{\beas}{\begin{eqnarray*}}
\newcommand{\eeas}{\end{eqnarray*}}
\newcommand{\bdm}{\begin{displaymath}}
\newcommand{\edm}{\end{displaymath}}
\newcommand{\ba}{\begin{array}}
\newcommand{\ea}{\end{array}}
\newcommand{\bi}{\begin{itemize}}
\newcommand{\ei}{\end{itemize}}
\newcommand{\ben}{\begin{enumerate}}
\newcommand{\een}{\end{enumerate}}
\newcommand{\bc}{\begin{center}}
\newcommand{\ec}{\end{center}}
\newcommand{\bfl}{\begin{flushleft}}
\newcommand{\efl}{\end{flushleft}}
\newcommand{\bfr}{\begin{flushright}}
\newcommand{\efr}{\end{flushright}}
\newcommand{\bd}{\begin{description}}
\newcommand{\ed}{\end{description}}
\newcommand{\bq}{\begin{quote}}
\newcommand{\eq}{\end{quote}}
\newcommand{\bfg}{\begin{figure}}
\newcommand{\efg}{\end{figure}}
\newcommand{\bt}{\begin{table}}
\newcommand{\et}{\end{table}}
\newcommand{\btb}{\begin{tabular}}
\newcommand{\etb}{\end{tabular}}
\newcommand{\btg}{\begin{tabbing}}
\newcommand{\etg}{\end{tabbing}}

\newcommand{\LLA}{\Large}

\newcommand{\NS}{\normalsize}
\newcommand{\SM}{\small}

\newcommand{\Ts}{\textstyle}

\newcommand{\e}{\enspace}

\newcommand{\itg}{\int \limits}

\newcommand{\eps}{\varepsilon}

\newcommand{\Li}{\mbox{Li}_{2}}

\newcommand{\lslash}
	   {\mbox{$ l \hspace{-0.9ex} \mbox{/} \hspace{-0.15ex} $}}

\newcommand{\qslash}
	   {\mbox{$ q \hspace{-1.1ex} \mbox{/} \hspace{-0.05ex} $}}

\newcommand{\Rfcs}{${\cal R}$ functions}
\newcommand{\Rfc}{${\cal R}$ function}

\begin{document}

\title{\bfr \NS MZ-TH 93-14 \\[1cm] \efr
\bf A new Method \\ for computing One-Loop Integrals \\[2cm]}
\author{L. Br\"ucher${}^{1}$,
J. Franzkowski${}^{2}$ \\[0.5cm]
{\NS Institut f\"ur Physik} \\
{\NS Johannes Gutenberg-Universit\"at} \\
{\NS Staudingerweg 7} \\
{\NS D-55099 Mainz} \\
{\NS Germany}
\and
D. Kreimer${}^{3}$ \\[0.5cm]
{\NS Dept.~of Physics} \\
{\NS Univ.~of Tasmania} \\
{\NS G.P.O.~Box 252C, Hobart 7001} \\
{\NS Australia} \\[1.5cm]}
\date{October 1993 \\[1cm]}

\maketitle

\begin{abstract}

We present a new program package for calculating one-loop Feynman
integrals, based on a new method avoiding Feynman parametrization and
the contraction due to Passarino and Veltman. The package is
calculating one-, two- and three-point functions both algebraically and
numerically to all tensor cases. This program is written as a package
for Maple. An additional Mathematica version is planned later.

\end{abstract}

\footnotetext[1]{e-mail: Bruecher@vipmza.physik.uni-mainz.de}
\footnotetext[2]{e-mail: Franzkowski@vipmza.physik.uni-mainz.de}
\footnotetext[3]{e-mail: Kreimer@physvax.phys.utas.edu.au}

\thispagestyle{empty}

\newpage

\section*{PROGRAM SUMMARY}

{\SM {\it Title of program: } ONELOOP.MA \\[0.4cm]
{\it Computer: } (i) VAX 4000-90 (ii) PC 486 DX \\[0.4cm]
{\it Operating system: } (i) VMS (ii) MS-DOS 6.0 \\[0.4cm]
{\it Programming language: } MAPLE V \\[0.4cm]
{\it Subprograms used: } CFCN.MA, R.MA, SIMPLE.MA, PV.MA \\[0.4cm]
{\it No. of lines in distributed program including subprograms: } 1650
\\[0.4cm]
{\it Keywords: } Electroweak theory, Feynman diagrams, one-loop corrections,
renormalization \\[0.4cm]
{\it Nature of physical problem: } The theoretical determination of cross
sections in particle processes requires the calculation of radiative
corrections. The most important contribution comes from the level of
one-loop Feynman diagrams which arise from the model under consideration,
usually the standard model of elementary particles \\[0.4cm]
{\it Method of solution: } This package is designed to evaluate
automatically one-loop integrals. It is making use of the properties of
${\cal R}$ functions, a class of special functions which simplifies the
evaluation of Feynman integrals \\[0.4cm]
{\it Typical running time: } The algebraic evaluation of a two-point
function up to the rank 2 tensor on a VAX 4000-90 needs less than 20s }

\section*{LONG WRITE-UP}

\section{Introduction}

In particle physics the calculation of one- and higher loop corrections
to particle processes is necessary to keep track with the increasing
accuracy of particle colliders. In this context it is very useful if a
fast and automatized calculation of the contributing Feynman diagramms
is provided by a computer. On the most basic loop correction order, the
one-loop level, several attempts for such programs are already
available, like FF \cite{Ff} -- calculating scalar one-loop integrals
in Fortran -- or FeynCalc \cite{Mer} -- a Mathematica \cite{Math}
package -- which exhibits the usual Passarino-Veltman coefficients of
tensor integrals in terms of scalar integrals. Both programs are using
the standard procedure for calculating one-loop diagrams, namely Feynman
parametrization, Wick rotation and the Passarino-Veltman contraction
for tensor structure (cf. \cite{Hoo} and \cite{Pas}).

Although this standard procedure became very popular during the last
decade, it turned out to be rather inconvenient for being implemented on
a computer if the integral involves higher tensor structure. For that
reason we investigated in a new method for calculating arbitrary
one-loop integrals which is already published for the two- and
three-point case (cf. \cite{Kr2} and \cite{Kr3}).

Our approach is completely different from the standard procedure and
the structure of other packages. We are not forced to invert big
matrices which is the most difficult step in computing the tensor
structure of the integrals. In our method the loop momentum
integrations are performed directly, using the residue theorem and
introducing \Rfcs \ (cf. \cite{Car}), a class of special functions with
some nice and useful features. Instead of big matrices we get some
recursion relations for the involved \Rfcs \ (cf. \cite{Kr1} or
\cite{Fra}), which allow a fast and memory efficient evaluation
of the integrals under consideration.

The program is designed to reduce the \Rfcs \ to a set of fundamental
functions which are rewritten in terms of logarithms and dilogarithms.
Up to now the different one-, two- and three-point functions are
programmed as procedures in Maple \cite{Map}. Therefore our results are
available numerically as well as algebraically. We tested our
procedures using Maple V on a VAX 4000-90 workstation.

\section{Description of the method}

A detailed description of the method may be found in \cite{Kr1} or
\cite{Fra}. Here we would like to give a brief introductory overview
only. The calculation is based on the well-known dimensional
regularization method, as described for example in \cite{Coll}.
Instead of introducing Feynman parameters the integral is
split up into orthogonal and parallel space (cf. \cite{Coll}),
so that for example the two-point integral takes on the form
\bea
\label{two}
\lefteqn{B^{(p_{0} p_{1})}(q, m_{1}, m_{2})} \\
& = & \frac{2\pi^{\frac{D-1}{2}}}{\Gamma(\frac{D-1}{2})}
\itg_{-\infty}^{\infty} \!\! dl_{\|} \itg_{0}^{\infty} \!\! dl_{\bot} \,
l_{\bot}^{D-2} \, \frac{(l_{\|})^{p_{0}} \, (l_{\bot})^{p_{1}}}
{[(l_{\|} + q)^{2} - l_{\bot}^{2} - m_{1}^{2} + i \varrho] \,
[l_{\|}^{2} - l_{\bot}^{2} - m_{2}^{2} + i \varrho]} \nonumber
\eea
Here we present the most general case, depending on one external
momentum $q$ and two arbitrary masses $m_{1}$ and $m_{2}$. The
splitting of space requires a distinction between $l_{\|}$ and
$l_{\bot}$ which we define as
\bea
l_{\|} & = & \frac{l\cdot q}{\sqrt{q^2}} \nonumber \\
\label{split1}
l_{\bot} & = & \sqrt{l_{\|}^2 - l^2}
\eea
In this notation $l_{\|}$ describes the component of $l$ which is parallel
to the external momentum $q$, whereas $l_{\bot}$ represents the orthogonal
complement.

The integrals (\ref{two}) can be reduced to a sum of \Rfcs \, by using
(\ref{IntForm}) with the help of partial fraction and residue theorem.
The complete procedure is described in detail in \cite{Fra} and \cite{Kr1}.
The result in the two-point-case may be written according to \cite{Kr2}:
\beas
\lefteqn{B^{(p_{0} p_{1})}(q, m_{1}, m_{2})} \\
& = & \frac{i \pi^{\frac{3}{2} - \eps}}{\Gamma(\frac{3}{2} - \eps)}
\, {\Ts \Gamma(\frac{1}{2} + \eps - \frac{p_{1}}{2}) \, \Gamma(\frac{3}{2} -
\eps + \frac{p_{1}}{2})} \, \frac{(-1)^{\Ts \frac{p_{1}}{2}} \,
(e^{- i \pi})^{-\eps}}{q \, (- 2 \eps + p_{1} + 1)} \\
& & \left[ (-1)^{p_{0} - [p_{0}]} \, {\Ts (\frac{q}{2} + A)^{1 +
p_{0} - [p_{0}]} \, B(\eps - \frac{p_{1} + [p_{0}]}{2}, \frac{[p_{0}] +
1}{2})} \right. \\
& & \hspace{1.9cm} {\cal R}_{- \eps + \frac{p_{1} + [p_{0}]}{2}} \left({\Ts -
\frac{p_{1} + 1}{2} + \eps, 1; - m_{1}^{2} + i \varrho, -
(\frac{q}{2} + A)^{2}} \right) \\
& & + \sum_{i=0}^{p_{0}} \, {p_{0} \choose i} \, (-q)^{p_{0}-i} \, {\Ts
(\frac{q}{2} - A)^{1 + i - [i]} \, B(\eps - \frac{p_{1} + [i]}{2},
\frac{[i] + 1}{2})} \\
& & \left. \hspace{3.6cm} {\cal R}_{- \eps + \frac{p_{1} + [i]}{2}}
\left({\Ts - \frac{p_{1} + 1}{2} + \eps, 1; - m_{2}^{2} + i \varrho, -
(\frac{q}{2} - A)^{2}} \right) \right]
\eeas
where $\eps = \frac{4 - D}{2}$ represents the usual dimensional
regularization parameter. We used the abbreviations
\beas
A & = & \frac{m_{2}^{2} - m_{1}^{2}}{2 q} \\[0.4cm]
[a] & = & \, \left\{ \e \ba{ll}
a, & \e \mbox{$a$ even} \\
a+1, & \e \mbox{$a$ odd}
\ea \right.
\eeas
The result of the three-point integral is obtained by a quite similar
procedure and may be found in \cite{Fra} or \cite{Kr3}.

\section{Evaluation and expansion of the \Rfcs}

After having expressed all possible integrals in terms of \Rfcs \ we
get easy and fast computable results by applying some recurrence
relations for the \Rfcs \ under consideration.

We achieve these relations by first reducing the number of different \Rfcs \
to a set of fundamental \Rfcs. This is possible due to the already
mentioned features of the \Rfcs, especially the relation (\ref{ParInc})
and (\ref{AssFk}). The reduction takes place in two steps. First of all
we increase the parameter of the \Rfc \ to get it close to 0 or 1.
This can easily be achieved by using formula (\ref{ParInc}). In the second
step we reduce the index of the \Rfc \ by virtue of a convenient formula
derived from (\ref{AssFk}). In the case of the two point-function it has
the form:
\bea
{\cal R}_t(b_1,b_2;z_1,z_2) & = & \frac{-1}{t+\beta-1} \, \lbrack \,
((t-1)z_1 z_2)
   \, {\cal R}_{t-2}(b_1,b_2;z_1,z_2) \\
   & & + \, \left( (1-t-b_1)z_1 + (1-t-b_2)z_2 \right)
   \, {\cal R}_{t-1}(b_1,b_2;z_1,z_2) \rbrack \nonumber
\eea
Now the power of this method is obvious. Both steps mentioned above may be
performed easily by computer, due to the recursive definition of the
formulae (\ref{ParInc}) and (\ref{AssFk}).

Up to now we have acquired a result which is exact to all powers of $\eps$,
though we are interested only in an expansion in powers of $\eps$.
Therefore we now substitute the expansion for the fundamental \Rfcs \
in terms of $\eps$, resulting in simply computable functions like
logarithms and dilogarithms. This is a rather cumbersome task but has to
be performed only once for each N-point function. The procedure, especially
its subtleties concerning analytical continuation, is described
in detail in another paper \cite{BFK}. For completeness we
here just cite the expansion for the scalar two-point case
\bea
\lefteqn{ {\cal R}_{-\eps}(-{\Ts \frac{1}{2}}+\eps,1,z_1,z_2)  =
   \frac{\pi i}{B(\eps,\frac{1}{2})} \,
       \frac{(z_2-z_1)^{\frac{1}{2}-\eps}}{z_2^{\frac{1}{2}}} } \nonumber \\
 & & +  {\Ts \frac{1}{2}} \, z_1^{-\eps} \, \left(\frac{z_1}{
  z_2}\right)^{1-\eps} \, \left[ \left(1+\sqrt{1-\frac{z_1}{z_2}}\right)^{-
1+2\eps} + \left(1-\sqrt{1-\frac{z_1}{z_2}}\right)^{-1+2\eps}
  \right] + {\cal O}(\eps^2) \nonumber
\eea

The three-point integrals
\beas
\lefteqn{C^{(p_{0} p_{1} p_{2})}(q_{1}, q_{2}, m_{1}, m_{2}, m_{3})} \\
& = & \frac{2\pi^{\frac{D-2}{2}}}{\Gamma(\frac{D-2}{2})}
\itg_{-\infty}^{\infty} \!\! dl_{0\|} \itg_{-\infty}^{\infty} \!\! dl_{1\|}
\itg_{0}^{\infty} \!\! dl_{\bot} \, l_{\bot}^{D-3} \,
\frac{(l_{0\|})^{p_{0}} \, (l_{1\|})^{p_{1}} \, (l_{\bot})^{p_{2}}}
{[(l_{0\|} + q_{1})^{2} - l_{1\|}^{2} - l_{\bot}^{2} - m_{1}^{2} + i
\varrho]} \\
& & \frac{1}{[(l_{0\|} + q_{20})^{2} - (l_{1\|} + q_{21})^{2} - l_{\bot}^{2} -
m_{2}^{2} + i \varrho] \, [l_{0\|}^{2} - l_{1\|}^{2} - l_{\bot}^{2} -
m_{3}^{2} + i \varrho]}
\eeas
are calculated in a quite similar way (cf. \cite{Fra} or \cite{Kr3}).
Again we treat the general case with two independent external momenta
$q_{1},q_{2}$ and three masses $m_{1},m_{2},m_{3}$. For the momentum
components we write in a quite natural generalization:
\bea
l_{0\|} & = & \frac{l\cdot q_1}{\sqrt{q_1^2}} \nonumber \\
l_{1\|} & = & - \frac{l\cdot {q'}_2}{\sqrt{{q'}_2^2}}\,;\quad {q'}_2 = q_2
	       - \frac{q_1\cdot q_2}{q_1^2}\,q_1 \nonumber \\[0.2cm]
\label{split2}
l_{\bot} & = & \sqrt{l_{0\|}^2 - l_{1\|}^2 - l^2} \\[0.2cm]
q_{20} & = & \frac{q_1\cdot q_2}{\sqrt{q_1^2}} \nonumber \\
q_{21} & = & \sqrt{q_{20}^2 - q_{2}^2} \nonumber
\eea
The evaluation is resulting in the fundamental \Rfcs \
\bea
\lefteqn{{\cal R}_{-2\eps}(\eps, \eps, 1; x, y, z) = z^{-2\eps} \, \left\{1
+ 2\eps^{2} \, \left[\Li \left(1 - \frac{x}{z}\right) + \Li \left(1 -
\frac{y}{z}\right) \right. \right.} \nonumber \\
& & + \ln \left(1 - \frac{x}{z}\right) \eta \left(x, \frac{1}{z}\right)
+ \ln \left(1 - \frac{y}{z}\right) \eta \left(y, \frac{1}{z}\right) +
\ln z \, \left[\eta\left(x - z,\frac{1}{1-z}\right) \right. \nonumber \\
& & \left. \left. \left. - \eta\left(x - z,-\frac{1}{z}\right) + \eta\left(y
- z,\frac{1}{1-z}\right) - \eta\left(y - z,-\frac{1}{z}\right)\right]\right]
\right\} + O(\eps^{3}) \nonumber \\
\lefteqn{{\cal R}_{1-2\eps}(\eps, \eps, 1; x, y, z) = (1 - 2 \eps) z^{1-2\eps}
+ \eps (x + y)} \nonumber \\
& & + 2\eps^{2} \left[- y \, \ln y - x \, \ln x + z \, \Li\left(1 -
\frac{x}{z}\right) + z \, \Li\left(1 - \frac{y}{z}\right)\right. \\
& & + z \, \ln \left(1 - \frac{x}{z}\right) \eta \left(x, \frac{1}{z}\right)
+ z \, \ln \left(1 - \frac{y}{z}\right) \eta \left(y, \frac{1}{z}\right) +
z \, \ln z \, \left[\eta\left(x - z,\frac{1}{1-z}\right) \right. \nonumber \\
& & \left. \left. - \eta\left(x - z,-\frac{1}{z}\right) + \eta\left(y
- z,\frac{1}{1-z}\right) - \eta\left(y - z,-\frac{1}{z}\right)\right]\right]
+ O(\eps^{3}) \nonumber
\eea
noting that our relations are valid in all kinematical regions, so that
the results of \cite{BFK} apply to all the cases of our program. For a
detailed discussion see \cite{BFK}.

\section{The one-loop functions}

Our aim is to calculate one-loop integrals both algebraically and
numerically. Therefore the algorithm is implemented in Maple \cite{Map},
a `language` for symbolic computations. To make the implementation for
any user most comfortable, the procedures are bound to a package, which
may be read from the Maple system during run-time. The user-interface
consists of three elementary functions
\begin{itemize}
 \item OneLoop1Pt$(p,m)$

 \item OneLoop2Pt$(p_0,p_1,q,m_1,m_2)$

 \item OneLoop3Pt$(p_0,p_1,p_2,q_{1},q_{20},q_{21},m_1,m_2,m_3)$
\end{itemize}
The functions directly correspond to the following integrals
\begin{itemize}
 \item One-point function:
\beas
A^{(p)}(m) & = & \int \! d^{D}l \, \frac{(l)^{p}}{[l^{2} - m^{2} +
i \varrho]}
\eeas
 \item Two-point function:
\beas
\lefteqn{B^{(p_{0} p_{1})}(q, m_{1}, m_{2})} \\
& = & \frac{2\pi^{\frac{D-1}{2}}}{\Gamma(\frac{D-1}{2})}
\itg_{-\infty}^{\infty} \!\! dl_{\|} \itg_{0}^{\infty} \!\! dl_{\bot} \,
l_{\bot}^{D-2} \, \frac{(l_{\|})^{p_{0}} \, (l_{\bot})^{p_{1}}}
{[(l_{\|} + q)^{2} - l_{\bot}^{2} - m_{1}^{2} + i \varrho] \,
[l_{\|}^{2} - l_{\bot}^{2} - m_{2}^{2} + i \varrho]}
\eeas
 \item Three-point function:
\beas
\lefteqn{C^{(p_{0} p_{1} p_{2})}(q_{1}, q_{2}, m_{1}, m_{2}, m_{3})} \\
& = & \frac{2\pi^{\frac{D-2}{2}}}{\Gamma(\frac{D-2}{2})}
\itg_{-\infty}^{\infty} \!\! dl_{0\|} \itg_{-\infty}^{\infty} \!\! dl_{1\|}
\itg_{0}^{\infty} \!\! dl_{\bot} \, l_{\bot}^{D-3} \,
\frac{(l_{0\|})^{p_{0}} \, (l_{1\|})^{p_{1}} \, (l_{\bot})^{p_{2}}}
{[(l_{0\|} + q_{1})^{2} - l_{1\|}^{2} - l_{\bot}^{2} - m_{1}^{2} + i
\varrho]} \\
& & \frac{1}{[(l_{0\|} + q_{20})^{2} - (l_{1\|} + q_{21})^{2} - l_{\bot}^{2} -
m_{2}^{2} + i \varrho] \, [l_{0\|}^{2} - l_{1\|}^{2} - l_{\bot}^{2} -
m_{3}^{2} + i \varrho]}
\eeas
\end{itemize}
Again we follow our notation which distinguishes between parallel and
orthogonal space. The masses are denoted by $m_{i}$ and the definitions
of (\ref{split1}) and (\ref{split2}) for the momenta are of course
still valid. We would like to emphasize the fact that in our notation
the indices $p_i$ represent the powers of the different components of
the loop momentum $l$ and should not be mixed with Lorentz indices of
the corresponding tensor. This notation is perhaps not very often used,
but it corresponds directly to our method of integration. The usual
notation may be recovered by an additional algorithm described in
section \ref{PV}.

The results of the one-loop functions are Laurent expansions in terms of
the ultraviolet regulator $\eps$. The output of the described procedures
consists of a list where the significant coefficients of this expansion are
denoted.
\\[0.4cm]
\bc
\btb{|l|l|}
\hline
input & output \\
\hline
OneLoop1Pt$(p,m)$ & $[\eps^{-1}\mbox{-{\em term}},\eps^0\mbox{-{\em term}}]$ \\
\hline
OneLoop2Pt$(p_0,p_1,q,m_1,m_2)$ & $[\eps^{-1}\mbox{-{\em term}},
\eps^0\mbox{-{\em term}}]$ \\
\hline
OneLoop3Pt$(p_0,p_1,p_2,q_{1},$ & $[\eps^{-2}\mbox{-{\em term}},
\eps^{-1}\mbox{-{\em term}},$ \\
\qquad \qquad $q_{20},q_{21},m_1,m_2,m_3)$ & \qquad \qquad $\eps^0\mbox{-{\em
term}},\{abb.\}]$ \\
\hline
\etb
\\[1cm]
\ec
In the output ``$\eps^{-1}$-term'' is meant to represent the divergent
part (the coefficient of $\frac{1}{\eps}$) whereas ``$\eps^{0}$-term''
is describing the finite part. Of course the $\eps^{-2}$-term has to
vanish in any case, but is left in order to check the accuracy of the
program in numerical calculations. In the three-point case ``abb.'' is
written for some abbreviations which are introduced to get a more
compact result. For the same reason we define a function $\Lambda$
which collects the ocurring $\eta$ functions in the three-point case:
\be
\Lambda(x,y) = \ln(y) \, \left[\eta\left(x-y,\frac{1}{1-y}\right) -
\eta\left(x-y,-\frac{1}{y}\right)\right] + \ln\left(1-\frac{x}{y}\right)
\, \eta\left(x,\frac{1}{y}\right)
\ee

Up to now the program is not regulating the infrared divergence of the
three-point function -- which may occur if one mass is set to 0 -- by
virtue of an infrared dimension parameter $\eps_{IR}$. Therefore the user
is expected to avoid this kind of divergence by introducing a (small) mass
parameter instead of 0.

Another difficulty arises if the external momenta are collinear, which
means that $q_{21} = 0$. Although this case is simpler to calculate,
because the parallel space is only one-dimensional, the program is not
taught to deal with this special case up to now. The user is asked to use a
small $q_{21}$-component instead.

As long as the arguments of the OneLoop functions are symbols the
output is given algebraically, whereas of course numbers inserted for
the arguments imply a numerical result. Moreover, in the numerical
case, the program sets the value of $\varrho$, the imaginary part of
the propagators, five digits higher than the numerical accuracy of the
whole calculation. In the algebraic case just 'rho' is returned.

For different calculations of the same OneLoop function it is of course
rather inconvenient to start the whole program again. Therefore our package
includes the functions OneLoopLib1Pt($n$), OneLoopLib2Pt($n$) and
OneLoopLib3Pt($n$) which generate a library of all OneLoop1Pt,
OneLoop2Pt and OneLoop3Pt functions respectively up to the tensor rank
$n$. Each function is written to one file. The filenames correspond to
the integrals in the following way:\\[1cm]
\btb{ll}
ONE0.RES     & OneLoop1Pt$(0,m)$ \\
TWO00.RES    & OneLoop2Pt$(0,0,q,m_1,m_2)$ \\
TWO10.RES    & OneLoop2Pt$(1,0,q,m_1,m_2)$ \\
THREE000.RES & OneLoop3Pt$(0,0,0,q_{1},q_{20},q_{21},m_1,m_2,m_3)$ \\
THREE100.RES & OneLoop3Pt$(1,0,0,q_{1},q_{20},q_{21},m_1,m_2,m_3)$ \\
THREE010.RES & OneLoop3Pt$(0,1,0,q_{1},q_{20},q_{21},m_1,m_2,m_3)$ \\
$\cdots$     & $\cdots$
\etb
\\[1cm]
To specify the path of the library the variable ``tensorpath'' may be
assigned to the desired directory.

\section{The Passarino-Veltman coefficients} \label{PV}

To receive a more familiar representation of the output a procedure is
added which extracts the Passarino-Veltmann coefficients from our
programs, PassVelt1Pt, PassVelt2Pt and PassVelt3Pt for the different
n-point cases. The following types of arguments -- here demonstrated for
instance for the two-point function -- are possible:
\begin{itemize}
\item PassVelt2Pt($n$): The tensor decomposition of the rank $M$ tensor
      two-point function
      \bdm
      \int \! d^Dl \, \frac{l_{\mu_1} \cdots l_{\mu_M}}{[(l + q)^{2} -
      m_{1}^{2} + i \varrho] \, [l^{2} - m_{2}^{2} + i \varrho]}
      \edm
      is given. The procedure returns a list consisting in the
      different coefficients, which are expressed
      in terms of the OneLoop tensor integrals, and the defining
      equation for the coefficients, for instance in the case $M=2$:
      \beas
      & & \left[C_{21} = - \frac{\mbox{OneLoop2Pt}(0,2)}{3-2\eps} \, , \,
      C_{20} = - \frac{\mbox{OneLoop2Pt}(0,2)}{q^2 \, (-3+2\eps)}
      \right. \\
      & & \left. + \frac{\mbox{OneLoop2Pt}(2,0)}{q^2} \, , \, C_{20}
      q_{\mu_1} q_{\mu_2} + C_{21} g_{\mu_1\mu_2} \right]
      \eeas
\item PassVelt2Pt($n$,full): The function returns the same as above, but
      inserts the results of the OneLoop tensor integrals explicitly.
\item PassVelt2Pt($n,q,m_{1},m_{2}$): The function returns the same as
      PassVelt2Pt($n$,full), but expressed in the user defined terms for
      $q,m_{1}$ and $m_{2}$, using not the default parameters. Of course
      numerical values are also possible.
\end{itemize}
In the three-point case a similar decomposition of the rank $M$ tensor
three-point function
\bdm
      \int \! d^Dl \, \frac{l_{\mu_1} \cdots l_{\mu_M}}{[(l + q_1)^{2} -
      m_{1}^{2} + i \varrho] \,[(l + q_2)^{2} -
      m_{2}^{2} + i \varrho] \, [l^{2} - m_{3}^{2} + i \varrho]}
\edm
is returned

If the variable ``tensorpath'' is set to some path, the PassVelt
procedure assumes that the OneLoop functions are written there as
described in the previous section. It then reads the necessary integrals
 from this directory. Otherwise, if ``tensorpath'' is not assigned,
PassVelt is forced to calculate the necessary OneLoop functions by its
own.

\section{An example}

To demonstrate how our package works we illustrate the calculation of the
electron self-energy in QED. Starting with QED Feynman rules we get:
\bea
\Sigma (q) & = & -ie^2 \mu^{4-D} \int \frac{d^Dl}{(2 \pi)^D} \, \gamma_{\mu} \,
   \frac{1}{\qslash + \lslash - m} \, \gamma_{\nu} \, \frac{g^{\mu \nu}}{l^2}
   \nonumber\\
 & = & -ie^2 \mu^{4-D} \int \frac{d^Dl}{(2 \pi)^D} \frac{\gamma_{\mu}
  (\qslash + \lslash + m) \gamma^{\mu} }{[ (l+q)^2 - m^2 ] \, l^2} \nonumber \\
 & = & -ie^2 \mu^{4-D} \int \frac{d^Dl}{(2 \pi)^D} \ \frac{1}{[(l+q)^2 - m^2
	] \, l^2}
	\,[ \gamma_{\mu} l_{\nu} \gamma^{\nu} \gamma^{\mu} +
	  \gamma_{\mu} q_{\nu} \gamma^{\nu} \gamma^{\mu} + m \gamma_{\mu}
	  \gamma^{\mu} ] \nonumber \\
 & = & -ie^2 \mu^{4-D} \int \frac{d^Dl}{(2 \pi)^D} \ \frac{1}{[(l+q)^2 - m^2
	] \, l^2}
	\, [ l_{\nu} \gamma^{\nu}(2-D) + q_{\nu} \gamma^{\nu}(2-D) + mD  ]
	\nonumber
\eea
Now we split into orthogonal and parallel space noting that the $\gamma$
matrix vector is split into $\gamma_{\|}$ and $\gamma_{\bot}$:
\bea
\Sigma (q) & = & -\frac{ie^2\mu^{4-D}}{(2 \pi)^D} \left\{ (2-D)\left[
   \gamma_{\|} \cdot \int d^Dl \ \frac{l_{\|}}{[(l+q)^2 - m^2] \, l^2} \right.
     \right. \\
    & & \left. - \gamma_{\bot} \cdot \int d^Dl \ \frac{l_\bot}{[(l+q)^2 - m^2
	] \, l^2} \right] \nonumber \\
   & & \left.+ [(2-D) q\cdot \gamma_{\|}  + D m] \int d^Dl \ \frac{1}
	{[(l+q)^2 - m^2] \, l^2} \right\} \nonumber \\
& = & -\frac{ie^2\mu^{4-D}}{(2 \pi)^D} \left\{(2-D)\left[ \gamma_{\|} \cdot
      B^{(10)}(q,m,0) - \gamma_{\bot} \cdot B^{(01)}(q,m,0) \right] \right.
       \nonumber \\
   & & \left. + [(2-D) q\cdot \gamma_{\|} + D m] \, B^{(00)}(q,m,0)
       \right\} \nonumber
\eea
At this stage of calculation we now take the opportunity to incorporate our
procedures which solve the ocurring integrals $B^{(p_0,p_1)}$. To be
specific we have to compute the functions OneLoop2Pt$(1,0,q,m,0)$,
OneLoop2Pt$(0,1,q,m,0)$ and OneLoop2Pt$(0,0,q,m,0)$:
\bea
\mbox{OneLoop2Pt}(1,0,q,m,0) \equiv B^{(10)}(q,m,0) & = & - \frac{i \pi^2}{2
	     \eps} \, q \ + \ \frac{\pi^2}{4q^3}\, \Bigg\{  \nonumber \\
\lefteqn{\hspace{-6 cm}
i\,\ln \left(1+ \frac {\sqrt {(q^2-m^2)^2 +4\,i\rho\,q^2}}
{q^2 + m^2} \right)
 \left[ m^4-q^4+(q^2-m^2)\,\sqrt {(q^2-m^2)^2 +4\,i\rho\,q^2}\, \right]
}\nonumber \\
\lefteqn{\hspace{-6 cm}
+i\,\ln \left(1- \frac {\sqrt {(q^2-m^2)^2 +4\,i\rho\,q^2}}
{q^2 + m^2} \right)
 \left[ m^4-q^4-(q^2-m^2)\,\sqrt {(q^2-m^2)^2 +4\,i\rho\,q^2}\, \right]
}\nonumber \\
\lefteqn{\hspace{-6 cm}
+i (m^4-q^4)\ln \left(-{\frac {\left (q^{2}+m^{2}\right )^{2}}{4\,q^{2}}}
\right)
+ 2\,i\ln (-m^2+i\rho)(q^4+m^2q^2-m^4)
}\nonumber \\
\lefteqn{\hspace{-6 cm}
+i\,(q^2-m^2)^2
\,\ln \left(-{\frac {\left (q^{2}-m^{2}\right )^{2}}{4\,q^{2}}}\right)
+2\,i (q^2-m^2)^2\ln(2)
}\nonumber \\
\lefteqn{\hspace{-6 cm}
-\pi \left(2q^4-(q^2-m^2) \sqrt{(q^2-m^2)^2 +4\,i\rho\,q^2}
\right)
-2\,iq^2(q^2-m^2)
}\nonumber \\
\lefteqn{\hspace{-6 cm}
+2\, i q^{4}\gamma+2\,iq^{4}\ln(\pi )
\Bigg\}
}\nonumber \\
\mbox{OneLoop2Pt}(0,1,q,m,0) \equiv B^{(01)}(q,m,0) & = & 0 \nonumber \\
\mbox{OneLoop2Pt}(0,0,q,m,0) \equiv B^{(00)}(q,m,0) & = & \frac{i \pi^2}{\eps}
 \ + \ \frac{\pi^{2}}{2 q^2}\, \Bigg\{ 2\, i \ln(2) (m^{2}-q^{2})
\nonumber \\
\lefteqn{\hspace{-6 cm}
+\, i\, \ln \left( 1- \frac {\sqrt {(q^2-m^2)^2 +4\,i\rho\,q^2}}
{q^2 + m^2}\right)\,\left(q^{2}+m^2 +\sqrt{(q^2-m^2)^2 +4\,i\rho\,q^2}\right)}
\nonumber \\
\lefteqn{\hspace{-6 cm}
+\,i\, \ln \left(1+ \frac {\sqrt {(q^2-m^2)^2 +4\,i\rho\,q^2}}
{q^2 + m^2} \right)\, \left(q^{2}+m^2 -\sqrt{(q^2-m^2)^2 +4\,i\rho\,q^2}\right)
} \nonumber \\
\lefteqn{\hspace{-6 cm}
+\,i\, \ln\left(-\frac{\left (q^{2}+m^{2}\right )^2}{4\,q^2} \right)
\,\left(q^2+ m^2 \right) -2\,i\ln(-m^{2}+i\rho)\,(q^{2}+m^{2})
} \nonumber \\
\lefteqn{\hspace{-6 cm}
+\,i\, \ln\left(-\frac{\left (q^{2}-m^{2}\right )^2}{4\,q^2} \right)
\,\left(-q^2+ m^2 \right)
+\, \pi \,\left(\sqrt{(q^2-m^2)^2 +4\,i\rho\,q^2}-2\,q^2 \right)
} \nonumber \\
\lefteqn{\hspace{-6 cm}
-2\,i\, q^2 \left(\, \ln(\pi )+\gamma_E -2 \right) \Bigg\}  } \nonumber
\eea
Inserting these functions and simplifying the expression one may rewrite
the result in the following way:
\bea
  \Sigma (q) & = & \frac{e^2}{16 \pi^2} \left\{(-q \cdot \gamma_{\|} + 4m)\,
         \left[ \frac{1}{\eps} + \ln 4\pi - \gamma + \ln \left(\frac{\mu^{2}}
     {m^{2}}\right) \right]\right. \nonumber \\
& & - q \cdot \gamma_{\|} \, \left[\left(\frac{m^{4}}{q^{4}}-1\right) \,
       \ln\left(1-\frac{q^{2}}{m^{2}}\right) + 1 + \frac{m^{2}}{q^{2}}
      \right] \nonumber \\
& & \left. -4m \left[\left(1-\frac{m^{2}}{q^{2}}\right)\, \ln\left(1-
     \frac{q^{2}}{m^{2}}\right) - \frac{3}{2}\right]\right\} + {\cal O(\eps)}
\\
& = & \frac{e^2}{16 \pi^2} \left\{(-\qslash + 4m)\,
         \left[ \frac{1}{\eps} + \ln 4\pi - \gamma + \ln \left(\frac{\mu^{2}}
     {m^{2}}\right) \right]\right. \nonumber \\
& & - \qslash \, \left[\left(\frac{m^{4}}{q^{4}}-1\right) \,
       \ln\left(1-\frac{q^{2}}{m^{2}}\right) + 1 + \frac{m^{2}}{q^{2}}
      \right] \nonumber \\
& & \left. -4m \left[\left(1-\frac{m^{2}}{q^{2}}\right)\, \ln\left(1-
     \frac{q^{2}}{m^{2}}\right) - \frac{3}{2}\right]\right\}
    + {\cal O(\eps)} \nonumber
\eea
which is the well-known result for the electron self-energy.

\section{Conclusion and further projects}

Our aim was to introduce a program package that gives any computer user
the possibilty to calculate one-loop particle processes without much
effort in manpower and hardware. At this stage of development the program
is understood to be the first step of a larger project. We do not claim
that the package is up to now completely bug-free and optimized. In fact we
are planning to incorporate the possibility to regulate infrared
divergences also in the dimensional regularization scheme and taking into
account higher powers of propagators as well as the special case of
collinear external momenta.

As a further project we plan to extend this package to the four- and
five-point function. Moreover we want to export our program package to
Mathematica \cite{Math}, because from our point of view Mathematica at
the moment is the more common language than Maple. Planned is also a
menu-driven user interface. But this might be rather difficult, because
we haven't yet found a convenient C-language package for algebraic
calculations. (For hints contact L. Br\"ucher)

\section*{Acknowledgements}

We would like to thank R. Stemler for his contribution in developing the
PassVelt procedures.

\newpage

\begin{appendix}
$\e$\\[0.5cm]
{\LLA \bf Appendix} \\

\setcounter{equation}{0}
\renewcommand{\theequation}{\mbox{A}.\arabic{equation}}

\subsection*{Conventions}

\begin{tabular}{ll}
   $D=4-2\eps$      &   dimension in the dimensional regularisation method \\
   $ {\cal R}_t(b;z)$  & \Rfc, t is called index, \\
		    &   $b=(b_1,\ldots,b_n)$ parameters, $z=(z_1,\ldots,z_n)$
			arguments \\
   $ \beta=\sum_{i=1}^n b_i$ & sum of parameters of an \Rfc      \\

\end{tabular}

\subsection*{Some fundamental properties of \Rfcs}

\begin{itemize}
\item formula for integrals
\begin{equation} \label{IntForm}
 \int \limits_0^{\infty} x^{\alpha-1} \, \prod_{i=1}^k (z_i+w_i x)^{-b_i}
  \, dx =
  B(\beta - \alpha,\alpha) \, {\cal R}_{\alpha - \beta}({\bf b},{\bf
  \frac{z}{w} })
  \, \prod_{i=1}^{k} w_i^{-b_i}
\end{equation}
\item increasing of parameters
\begin{equation} \label{ParInc}
 \beta\,{\cal R}_t({\bf b},{\bf z})\ =\
	(\beta + t)\, {\cal R}_t({\bf b}+e_i,{\bf z})
	   \ -\ tz_i\, {\cal R}_{t-1}({\bf b}+e_i,{\bf z})
\end{equation}
\item associated functions (k is the number of parameters)
\begin{equation}       \label{AssFk}
 \sum_{i=0}^k \ A_i^{(k)}(t,{\bf b},{\bf z})
	    {\cal R}_{-t-i}({\bf b},{\bf z})\ =\ 0
\end{equation}
with
\be                 \label{ADef}
 A_i^{(k)}(t,{\bf b},{\bf z})= \frac{(t,i)(\beta-t-k,k-i)}{t(\beta-t-k)}
  (t+i- \sum_{j=1}^k b_j z_j  \frac{\partial}{\partial z_j} ) E_i^{(k)}(z)
\ee
$E_i^{(k)} (z) $ is the elementary symmetric function, gained from
\be                \label{elemsym}
 (1+xz_1) \cdots (1+xz_k) = \sum_{i=0}^k x^i E_i^{(k)} (z)
\ee

\end{itemize}

\end{appendix}

\end{document}